\documentstyle[seceq,preprint,epsf]{jpsj}
\title
{Microscopic Study of Quantum Vortex-Glass Transition 
Field in Two-Dimensional Superconductors}
\author{Hideharu Ishida, Hiroto Adachi, and Ryusuke Ikeda}
\inst{Department of Physics, Kyoto University, Kyoto 606-8502}
\recdate{June 4, 2001}

\abst{The position $B_{\rm vg}$ of a field-tuned superconductor-insulator 
quantum transition occuring in disordered thin films is examined within the 
mean field approximation and starting from a hamiltonian of BCS type (favoring 
the $s$-wave pairing) with a random potential, and effects of an 
electron-electron repulsion on the transition field $B_{\rm vg}$ are also 
considered. Our calculation shows that the (microscopic) 
disorder-induced reduction of $B_{\rm vg}$ suggested experimentally cannot be 
explained without taking account of the familiar interplay between 
the randomness and the electron-electron repulsion enhancing 
the {\it quantum} superconducting fluctuation in such systems. 
}

\kword{superconductor-insulator transition, vortex-glass transition,
superconducting fluctuation}

\begin{document}
\newcommand{\r}{{\bf r}}
\newcommand{\A}{{\bf A}}
\newcommand{\q}{{\bf q}}
\newcommand{\Q}{{\bf Q}}
\newcommand{\p}{{\bf p}}
\newcommand{\k}{{\bf k}}

\sloppy
\maketitle

\section{Introduction}

There have been many reports\cite{ex1,ex2,ex3} on resistive data 
suggestive of a strong 
quantum fluctuation effect in disordered thin films under magnetic fields 
perpendicular to the film plane. In such systems, the temperature dependence 
of the sheet resistance $R(T)$ in higher fields than a field $B_c$ 
is typically insulating, while in lower fields 
it decreases on cooling, in most cases, in a manner 
suggesting the presence of a magnetic field-tuned superconducting
transition at zero temperature ($T=0$). It was found 
experimentally\cite{ex2,ex3} that $B_c$ decreases with decreasing the film 
thickness or increasing the impurity density, both of 
which enhance the high temperature sheet resistance $R_r$. 

Fisher\cite{MPAF} has proposed by extending the theoretical argument for 
zero field case\cite{FGG} to nonzero fields ($B \neq 0$) that, 
at a field $B_{\rm vg}$ implying such a 
quantum superconducting transition point to be identified with
$B_c$, the resistance curve $R(T)$ at 
lower temperatures should be independent of $T$ and take a universal 
constant value. However, extending directly the description\cite{FGG} 
in zero field to $B \neq 0$ case 
is theoretically invalid because 
a superconducting phase in the presence of the
field-induced vortices is not the Meissner phase but a kind of a vortex
glass (VG) phase\cite{DF,RI1} induced by a pinning disorder and with {\it no} 
Meissner effect. On the other hand, 
one of the present authors has pointed out\cite{RI2} 
that, in the  
field range where the pinning disorder is negligible, the 
insulating behavior of resistance in $T \neq 0$ 
can intrinsically arise due to the quantum 
superconducting fluctuation, and that, if the normal contribution
$\sigma_n$ to the conductivity is negligible, a flat $R(T)$ curve with 
approximately the 
value of quantum resistance $R_Q=\pi \hbar /2 e^2 = 6.5({\rm k}\Omega)$ 
is expected along the quantum melting (crossover) line at low temperatures 
which, as far as the ordinary dirty limit is valid, is insensitive 
to $T$. Also, it has been recently\cite{RI1} pointed out 
in terms of a phenomenological Ginzburg-Landau (GL) action 
with pinning disorder terms that, irrespective of the details of
dynamics of the superconducting fluctuation, the disorder-induced contribution 
to the $T=0$ conductance at $B_{\rm vg}$ should be, in contrast to the original argument,\cite{MPAF} a {\it nonuniversal} constant 
dependent on the strengths of pinning and fluctuation. 

However, in the present problem including a static disorder on the 
{\it electronic} level, a simple analysis starting from a GL action is 
insufficient, and more or less, 
one needs to return to a microscopic electronic hamiltonian. 
In a previous work,\cite{RI3} we have found, by deriving a quantum GL action, 
a significant enhancement of quantum fluctuation in nonzero fields 
due to an interplay between the microscopic disorder and a repulsive 
interaction between electrons. 
However, the familiar derivation of a GL action done there has neglected 
spatial variations of the coefficient of each term in the resulting GL model. 
A model of such spatial variations used widely in a GL theory 
is a random potential {\it for 
the pair-field} implying\cite{RI1,Lub} a randomness of $T_{c0}$, where $T_{c0}$ is the mean field transition temperature in zero field and in clean limit. 
To clarify whether the idea\cite{RI1,RI2} 
based on the quantum superconducting fluctuation 
is applicable to real systems, we have to take account of such a GL random 
potential term which induces, in a finite field, a pinning 
disorder\cite{LO} for the 
vortices. A crucial point is that, in the present issue, such a GL
random potential arises from the {\it microscopic} disorder $u({\bf r})$ 
which {\it simultaneously} enhances the (quantum) superconducting 
fluctuation. Note that, for a decrease and vanishing of resistivity in
$B \neq 0$, the superconducting fluctuation effect and the
vortex pinning are competitive with each other. Due to this competition 
originating from the same microscopic disorder, the disorder-induced 
decrease\cite{ex1,ex2} of $B_{\rm vg}$ may not be necessarily expected 
theoretically. The purpose of the present study is to give an answer to 
this question by starting from a microscopic model. 

To examine the quantum superconducting fluctuation\cite{RI2} and the
vortex glass fluctuation\cite{RI1}, arising from the pinning disorder, on
the same footing, one needs an appropriate quantum GL action with 
spatially-dependent coefficients. Within the approximation taking account 
only of the pair-field (superconducting order parameter) $\Psi$ belonging 
to the lowest Landau level (LLL), the quantum GL action is phenomenologically 
expected to take the following form
$$
S_{\rm ran}= \int d^2r 
\biggl[ \, \beta \sum_\omega (\mu(0) + \gamma|\omega|) \, 
|\Psi_\omega(\r)|^2 + \int du \biggl( \delta \mu(\r) |\Psi(\r,u)|^2 +
\frac{U_4}{2} |\Psi(\r,u)|^4 \biggr) \biggr], \eqno(1.1)
$$
where $u$ is an imaginary time, $\omega$ a Matsubara frequency, 
$\delta \mu({\bf r})$ a static random potential implying a spatial variation
of $T_{c0}$, $\Psi(\r,u)$ ($= N^{1/2}(0) 
<\psi_\uparrow(\r,u)\psi_\downarrow(\r,u)>$) 
$= \sum_\omega \Psi_\omega(\r) \exp(-{\rm i} \omega u)$, and 
$\gamma$ measures\cite{RI3} 
the dissipation strength and was denoted as $\gamma_1$ in ref.9. Since we 
focused on the LLL modes, eigenvalues of gradient terms were already 
absorbed into $\mu(0)$. 
For simplicity, a randomness of the quadratic term was taken in eq.(1.1) in a 
local form, i.e., up to the lowest order in the gradient. In $\S 2$, an 
action corresponding to a replicated version of the action (1.1) will be 
microscopically derived by neglecting the electron 
repulsion and focusing mainly on the low $T$ and high $B$ region. 
In $\S 3$, our formulation for determining the {\it Gaussian}
vortex-glass transition field $B_{{\rm vg},0}$ at $T=0$ will 
be explained and is applied both to the ordinary dirty limit with no 
repulsive interaction and to the case with an electron-repulsion. We find 
that, in contrast to the experimental observation,\cite{ex1,ex2,ex3} $B_{{\rm vg},0}$ in the ordinary dirty limit increases with increasing the (microscopic)
disorder, mainly reflecting an enhancement of the corresponding $H_{c2}(0)$ ($= H_{c2}^d(0) \propto T_{c0}/\tau$). Further, the analysis is extended 
to the case with a repulsive interaction, and we find that the 
quantum LLL fluctuation and a decrease of $H_{c2}(0)$ due to an 
inclusion of an electron-repulsion are origins of a reduction of $B_{\rm vg}$ 
consistent with experimental observation. An attempt of computation on 
low $T$ behaviors of $\gamma$ is given in $\S 4$ to demonstrate the presence 
of a region in which the GL coefficients are insensitive to $T$. 
Our results are summarized in $\S 5$ together with a comment. 

\section{Model and Derivation of Pinning Vertex}

As a microscopic basis of our analysis, we first start from the $s$-wave 
BCS Hamiltonian with random potential in two dimensional (2D) case 
$$ H=\int_{\bf r} \biggl[ \sum_\sigma \psi_\sigma^\dagger(\r)\bigg(
\frac{\hbar^2}{2m} 
(-{\rm i}\nabla-\frac{\pi}{\phi_0}\A(\r))^2+u(\r)\bigg)\psi_\sigma(\r) 
+ g_{\rm BCS} \, \psi_\uparrow^\dagger(\r)\psi_\downarrow^\dagger(\r) 
\psi_\downarrow(\r)\psi_\uparrow(\r) \biggr]. \eqno(2.1)
$$
For the moment, we will not take account of electron-electron repulsions and 
postpone its inclusion to $\S 3$. In 
eq.(2.1), $g_{\rm BCS} < 0$, $\phi_0$ is the flux quantum for Cooper pairs, 
$\psi_\sigma$ is an annihilation 
operator of electron with spin $\sigma$ ($=\uparrow$, 
$\downarrow$), and the applied field ${\bf B}= {\rm curl} {\bf A}$ is 
perpendicular to the 2 D plane. The random potential $u(\r)$ has zero mean and 
the Gaussian ensemble ${\overline {u(\r) u(\r')}} = (2\pi N(0) \tau)^{-1} 
\delta^{(2)}(\r-\r')$, 
where $\tau$ is the elastic relaxation time, $N(0)$ the density of states 
at Fermi surface, and the overbar denotes the microscopic random average. Following the study of zero field transition point in the literature\cite{Fin}, the dependence of physical quantities on 
the film thickness is assumed to arise entirely through the repulsive 
interaction effect. In most part of this paper, our analysis is based on the 
ordinary dirty limit in 2D (with no interaction effect) 
in which the disorder strength is measured by 
$(k_F l)^{-1} = \tau/(2 \pi N(0) l^2)$.  

In our microscopic analysis which follows, the so-called quasi-classical 
approximation will be used. In this treatment, a gauge-invariant gradient 
${\bf Q} = -{\rm i} \nabla + 2 \pi {\bf A}/\phi_0$ operating on a $\Psi$ is 
regarded as a wavevector (c-number) in the electronic process. 
Further, before focusing on the LLL modes of $\Psi$, the coefficients of 
the quadratic and quartic terms in the action are found to be functions 
of ${\bf Q}^2$, while ${\bf Q}^2$ is transformed, after 
operating on a LLL $\Psi$ mode, into $r_B^{-2} \equiv 2 \pi B/\phi_0$ where 
$B$ is the uniform flux density. Hence, after this transformation ${\bf Q}^2 
\to r_B^{-2}$, the coefficient of each term in the action, except a term 
leading to the vortex pinning, may be 
written as a constant. In addtion, we note regarding the validity of the 
LLL approximation that, near $T=0$, the field range over which the LLL 
approximation for the pair field is valid is quite broad even below 
$H_{c2}(0)$ according to ref.8. Under these conditions 
and within the ordinary dirty limit neglecting the localization effect and 
the electron-electron repulsion, the coefficients $\mu(0)$, 
$\gamma$, and $U_4$ in the resulting eq.(1.1) are well-known and given 
by\cite{RI3,Wert} 
$$
U_4^{(0)} = {{8 \pi \tau^3} \over {\beta N(0)}} 
\sum_{\epsilon>0} ( \, \Gamma(2\epsilon; B) \, )^3, \eqno(2.2)$$ 
$$\mu^{(0)}(0) = (N(0) |g_{\rm BCS}|)^{-1} - 4 \pi \tau \beta^{-1} 
\sum_{\epsilon>0} \Gamma(2\epsilon; B), \eqno(2.3)$$
$$\gamma^{(0)} = 4 \pi \beta^{-1} \tau^2 
\sum_{\epsilon>0} ( \, \Gamma(2\epsilon; B) \, )^2, \eqno(2.4)$$
where $\Gamma(2\epsilon; B)=(2|\epsilon|+2 \pi B D/\phi_0)^{-1} \tau^{-1}$ implies $\Gamma_{\bf Q}(2\epsilon)$ with $Q^2 = 2 \pi B/\phi_0$, 
$\Gamma_{\bf q}(\epsilon+\epsilon')=\tau^{-1}(|\epsilon+\epsilon'| + D q^2)^{-1}$ is the diffusion propagator with momentum ${\bf q}$, $\epsilon$ and $\epsilon'$ are Matsubara frequencies for fermions, 
and the diffusion constant $D$ is given in terms of the mean free 
path $l=k_F \tau/m$ by $D=l^2/(2 \tau)$. Through this paper, 
the (impurity-averaged) electron propagator is, as usual, given by 
$G_\k(\epsilon)=({\rm i}\epsilon - \xi_k +({\rm i}/2\tau){\rm sgn}
(\epsilon))^{-1}$. 

Next, to explain how the $\delta \mu({\bf r})$ term, neglected in previous 
studies,\cite{RI3} 
appears in the present formulation, let us first imagine the 
phenomenological action (1.1) to be rewritten in a replicated form. 
Following the standard treatment for an action of GL type, the 
replicated form\cite{Lub} of eq.(1.1) will be given by
$$S^n_{\rm ran} = \sum_{\alpha=1}^n \biggl[ \int_{\bf r} \beta \sum_\omega (\mu(0) + \gamma|\omega|) \, |\Psi^{(\alpha)}_\omega(\r)|^2 
+ \frac{U_4}{2} \int du \int_{\bf k} |\rho^{(\alpha)}_{\bf k}(u)|^2 $$
$$- \sum_{\alpha'=1}^n  
\int du_1 \int du_2 \int_{\bf k} {{U_p({\bf k})} \over 2} \, 
\rho^{(\alpha)}_{\bf k}(u_1) \rho^{(\alpha')}_{-{\bf k}}
(u_2) \biggr], \eqno(2.5)$$
where $\alpha$ and $\alpha'$ denote the replica indices, $u_1$ and $u_2$ are 
imaginary times, and the bare (${\bf k}$-dependent) pinning vertex function 
$U_p({\bf k})$ and $\rho^{(\alpha)}_{\bf k}(u)$ are Fourier-transformations, 
respectively, of ${\overline {\delta \mu(\r) \, \delta 
\mu(\r')}}$ and of $|\Psi^{(\alpha)}({\bf r}, u)|^2$. As seen below, the ${\bf k}$-dependence of the bare pinning vertex $U_p({\bf k})$ is characteristic of the high $B$ and low $T$ 
region. 

A microscopic derivation of the last term of eq.(2.5) will be explained here. 
Within the quasiclassical approximation\cite{Wert}, 
the third (replica off-diagonal) 
term of eq.(2.5) generally takes the following form 
$$\Delta S^n_{\rm ran} = - {{\beta^2} \over 2} \sum_{\alpha=1}^n
\sum_{\alpha'=1}^n \int d^2r \sum_\omega \sum_{\omega'} {\cal F}([{\bf Q}_s]; 
\, \omega, \omega') $$
$$\times (\Psi^{(\alpha)}_\omega({\bf r}_1))^*  \, \Psi^{(\alpha)}_\omega({\bf
r}_2) \, (\Psi^{(\alpha')}_{\omega'}({\bf r}_3))^*  \,
\Psi^{(\alpha')}_{\omega'}({\bf r}_4) \biggr|_{{\bf r}_1, {\bf r}_2, {\bf r}_3,
{\bf r}_4  \to {\bf r}}, \eqno(2.6)$$
where ${\bf Q}_s = -{\rm i}\nabla_s + (2 \pi/\phi_0){\bf A}({\bf r}_s)$, 
${\cal F}([{\bf Q}_s])$ is a bare vertex of the quartic 
term induced by the random-average, and $[{\bf Q}_s]$ implies dependences 
on ${\bf Q}_1^*$, ${\bf Q}_2$, ${\bf Q}_3^*$, and ${\bf Q}_4$. Eq.(2.6) 
includes terms off-diagonal in the replica indices as a reflection of the fact 
that no impurity lines carry finite 
frequencies. Due to the use of the quasiclassical approximation, the
spatial nonlocality, arising from the nonzero magnetic field, in
${\cal F}(\cdot \cdot \cdot)$ appears only in the form of the gauge-invariant
gradients ${\bf Q}_s$ operating the pair-fields $\Psi({\bf r}_s)$. Hereafter, 
frequency dependences of ${\cal F}$ will be neglected. 

Below, let us see how $\Delta S_{ran}^n$ will lead to the form 
of the last term 
of eq.(2.5) in a consistent approximation with the derivations of $U_4^{(0)}$, 
$\mu^{(0)}(0)$, and $\gamma^{(0)}$. 
Namely, as in the derivation of $U_4^{(0)}$, the diagrams expressing 
${\cal F}$ are selected under the condition, equivalent to the neglect of 
localization effect of noninteracting electrons, that the impurity lines, each 
of which carries the factor $(2 \pi N(0) \tau)^{-1}$, 
do not cross to each other in the diagrams. 
The diagrams thus obtained expressing ${\cal F}$ are the same as those 
in ref.14, where the diagrams were examined from the viewpoint of mesoscopic 
fluctuation, and their examples belonging to the same family are given 
in Fig.1. Effects of an electron-repulsion on $\Delta S_{ran}^n$ 
will not be examined here. We simply note that, just like those in $U_4$, each 
term perturbative in a short-ranged repulsive interaction in 
the function ${\cal F}([{\bf Q}_s])$ is convergent in $T \to 0$ limit. 
The first three diagrams 
of those in Fig.1 are given in Fig.2, on which we will 
focus here to clarify key features common to all diagrams in Fig.1. In the figures, the single solid curves denote $G_k(\epsilon)$'s, four double solid lines denote $\Psi({\bf r}_i)$ or $\Psi^*({\bf r}_i)$, and ${\bf Q}_i$ (${\bf Q}^*_i$) operates on $\Psi({\bf r}_i)$ ($\Psi^*({\bf r}_i)$). The sum of diagrams in 
Fig.2 contributes to ${\cal F}([{\bf Q}_s])$ in the manner 
$${\cal F}^{(2)} ([{\bf Q}_s]) = (N(0) \beta)^{-2} 
\sum_{\epsilon > 0} \sum_{\epsilon' > 0} \Gamma_{{\bf Q}^*_1}(2\epsilon) 
\Gamma_{{\bf Q}_2}(2\epsilon) \Gamma_{{\bf Q}^*_3}(2\epsilon') 
\Gamma_{{\bf Q}_4}(2\epsilon') \, (2\pi N(0) \tau)^{-2} \int_{\bf q} 
\Gamma_{{\bf q} + {\bf Q}_3^*- {\bf Q}_2}(\epsilon+\epsilon') $$ $$\times 
\Gamma_{\bf q}
(\epsilon+\epsilon') \int_{\bf k} G_{{\bf k}+\Q_2}(\epsilon) 
G_{\bf k}(-\epsilon) G_{{\bf k}+\Q^*_3+{\bf q}}(-\epsilon') 
G_{{\bf k}+{\bf q}}(\epsilon') \, 
(I^{(2a)} + I^{(2b)} + I^{(2c)}), \eqno(2.7)$$
where 
$$I^{(2a)} = \int_{\bf k} G_{\k+\Q^*_1}(\epsilon) G_{\k}(-\epsilon) 
G_{\k+{\bf q}+\Q_4}(-\epsilon') G_{\k+{\bf q}}(\epsilon'), $$
$$I^{(2b)} = - (2\pi \tau N(0))^{-1} \int_{{\bf k}_1} G_{{\bf k}_1+{\bf Q}^*_1}(\epsilon) G_{{\bf k}_1}(-\epsilon) G_{{\bf k}_1+{\bf q}}(\epsilon') \int_{{\bf k}_2} G_{{\bf k}_2+{\bf Q}_1^*}(\epsilon) G_{{\bf k}_2+{\bf q}+{\bf Q}_4}(-\epsilon') G_{{\bf k}_2+{\bf q}}(\epsilon'), $$
and $$I^{(2c)} =  
- (2\pi \tau N(0))^{-1} \int_{{\bf k}_1} G_{{\bf k}_1+{\bf Q}^*_1}(\epsilon) G_{{\bf k}_1}(-\epsilon) G_{{\bf k}_1+{\bf q}+{\bf Q}_4}(-\epsilon') \int_{{\bf k}_2} G_{{\bf k}_2}(-\epsilon) G_{{\bf k}_2+{\bf q}+{\bf Q}_4}(-\epsilon') G_{{\bf k}_2+{\bf q}}(\epsilon'), \eqno(2.8)$$
and $I^{(2a)}$ is the contribution of Fig.2(a) and so forth. However, a cancellation occurs between the three diagrams in the manner that 
the sum $I^{(2a)} + I^{(2b)} + I^{(2)}$ reduces to $2 \pi N(0) \tau^4 (2\epsilon+2\epsilon'+D(Q_1^*)^2 + DQ_4^2 + 2 D(q^2 + {\bf q}\cdot({\bf Q}_4-{\bf Q}_1^*)))$. It is easily understood that one more cancellation similar to this arises 
when all diagrams of Fig.1 are summed up, and the contribution of Fig.1 to ${\cal F}$ becomes 
$${\cal F}^{(1)}({\bf Q}, \Delta{\bf Q}) 
= \biggl( {{2 \tau^2 } \over { \beta N(0)}} \biggr)^2 
\sum_{\epsilon>0} \sum_{\epsilon'>0} (\, \Gamma_{\bf Q}(2\epsilon) 
\Gamma_{\bf Q}(2\epsilon') \, )^2 \int_{\bf q} \Gamma_{{\bf q}+\Delta{\bf Q}}(\epsilon+\epsilon') \, \Gamma_{\bf q}(\epsilon+\epsilon') $$ $$\times (\epsilon+\epsilon'+DQ^2 + Dq^2 + D{\bf q}\cdot\Delta{\bf Q})^2 \tau^2, \eqno(2.9)$$
where $\Delta{\bf Q}={\bf Q}_4-{\bf Q}_1^*={\bf Q}_3^* - {\bf Q}_2$, 
and all ${\bf Q}_s^2$ and $({\bf Q}_s^*)^2$ were expressed as ${\bf Q}^2$, 
because the external pair-fields are assumed to be in LLL. Further, using the 
fact that, after operating a LLL eigenfunction, ${\bf Q}^2$ changes into 
$r_B^{-2}=2 \pi B/\phi_0$, eq.(2.9) implies that ${\cal F}^{(1)}$ takes 
the form 
${\cal F}^{(1)}(\Delta{\bf Q}) = U_p f^{(1)}( \, t; \, r_B \Delta{\bf Q} 
\, )$, where 
$$U_p = \biggl({4 \over \pi} {{r_B  \, \tau} \over 
{N(0) l^2} } \biggr)^2, \eqno(2.10)$$
and $t=8 \pi \tau/(\beta l^2 r_B^{-2})$. Although other families of diagrams 
are also found to have similar structures to ${\cal F}^{(1)}$, the full expression ${\cal F}$ obtained after summing them up has a highly complicated 
$\Delta{\bf Q}$-dependence. Here we merely mention that, if formally setting 
$\Delta{\bf Q}=0$, the full ${\cal F}$ becomes the simplified form: 
$${\cal F}(\Delta{\bf Q}=0) \simeq 12 \biggl({{\tau^2} \over {N(0) \beta}} 
\biggr)^2 \sum_{\epsilon>0} \sum_{\epsilon'>0} \Gamma_{\bf Q}(2\epsilon) \Gamma_{\bf Q}(2\epsilon') \int_{\bf q} (\Gamma_{\bf q}(\epsilon+\epsilon'))^2, 
\eqno(2.11)$$ 
which is estimated as $0.14 U_p$ in low $t$ limit. 
Since, as is seen below, it is difficult to write down a 
concrete expression of $U_p({\bf k})$ resulting from the full 
${\cal F}(\Delta{\bf Q})$, we will merely explain below some 
properties of the full $U_p({\bf k})$. 

At this stage, let us rewrite the terms of eq.(2.5) entirely in terms of the 
LLL fluctuation field $\varphi^{(\alpha)}_0(p,\omega)$ so that $\rho^{(\alpha)}_{\bf k}(u)=\sum_\omega \rho^{(\alpha)}({\bf k}, \omega) e^{-{\rm i}\omega u}$ 
is 
expressed in a Landau gauge, by 
$$\rho^{(\alpha)}({\bf k}, \Omega) = \sum_p 
\exp(-{{{\bf k}^2 r_B^2} \over 4} + {\rm i}p k_x 
r_B^2) \sum_{\omega} \, (\varphi^{(\alpha)}_0(p+k_y/2, \omega+\Omega))^* 
\, \varphi^{(\alpha)}_0(p-k_y/2, \omega), \eqno(2.12)$$
where $\varphi_0^{(\alpha)}(p,\omega)$ is defined as $\Psi_\omega({\bf r}) = \sum_p \varphi_0(p, \omega) u_{0, p}({\bf r})$ in terms of a LLL eigen 
function $u_{0, p}$. If the $\Delta {\bf Q}$-dependences in ${\cal F}$ are 
neglected, the resulting pining vertex is local (i.e., the function $U_p({\bf k})$ is ${\bf k}$-independent). This approximation is valid in high $T$ and low $B$ region defined by $t \gg 1$, i.e., $T \gg 0.15 T_{c0} B/H_{c2}^d(0)$. In contrast, at low $T$ and high $B$ of our interest, any microscopic length 
on the pair-fields is scaled by $r_B$, and thus, the $\Delta{\bf Q}$-dependences are no longer negligible. Actually, we have the relations in a Landau gauge 
$$\int_{\bf r} \Delta{\bf Q} 
(\Psi^{(\alpha)}({\bf r}_1))^* \Psi^{(\alpha)}({\bf r}_2) (\Psi^{(\alpha')}({\bf r}_3))^* \Psi^{(\alpha')}({\bf r}_4)|_{{\bf r}_1, {\bf r}_2, {\bf r}_3, {\bf r}_4 \to {\bf r}} $$
$$={\rm i} \sum_{p, p'} \int_{\bf k} ({\hat z} \times {\bf k}) \, 
v_{\bf k} e^{{\rm i}(p-p')k_x r_B^2} 
(\varphi_0^{(\alpha)}(p-k_y/2))^* \varphi^{(\alpha)}_0(p+k_y/2) (\varphi^{(\alpha')}_0(p'+k_y/2))^* \varphi^{(\alpha')}_0(p'-k_y/2) ,$$ 
$$\int_{\bf r} (\Delta {\bf Q})^2 
(\Psi^{(\alpha)}({\bf r}_1))^* \Psi^{(\alpha)}({\bf r}_2) (\Psi^{(\alpha')}({\bf r}_3))^* \Psi^{(\alpha')}({\bf r}_4)|_{{\bf r}_1, {\bf r}_2, {\bf r}_3, {\bf r}_4 \to {\bf r}} $$ $$=2 \sum_{p, p'} \int_{\bf k} \biggl( r_B^{-2} - {{\bf k^2} \over 2} \biggr) \, v_{\bf k} e^{{\rm i}(p-p')k_x r_B^2} 
(\varphi_0^{(\alpha)}(p-k_y/2))^* \varphi^{(\alpha)}_0(p+k_y/2) (\varphi^{(\alpha')}_0(p'+k_y/2))^* \varphi^{(\alpha')}_0(p'-k_y/2) \eqno(2.13)$$ 
and so on, where $v_{\bf q}=\exp(-{\bf q}^2 r_B^2/2)$. Namely, any 
$r_B \Delta{\bf Q}$-dependence changes, after the Landau level representation, 
into a $r_B {\bf k}$-dependence or a constant. Hence, $U_p({\bf k})$ in eq.(2.5) generally has the form 
$$U_p({\bf k}) = U_p \, f_{00}(t; \, {\bf k} r_B ). \eqno(2.14)$$ 
If the Fermi surface is isotropic, $f_{00}$ is a function of $k^2r_B^2$. Below, for brevity, we will focus on this isotropic case. At low temperatures 
$t \ll 1$, $f_{00}$ is not sensitive to the material parameters 
included in the definition of $t$ but depends merely on $k^2r_B^2$. 
Unfortunately, it is difficult to find an exact form of such 
$k$-dependences in terms of eq.(2.13) which implies that a 
$\Delta {\bf Q}$-expansion does not reduce to a systematic 
${\bf k}$-expansion. Nevertheless, we have tried to estimate 
the $k$-dependence of $f_{00}$ at $T=0$ occuring from the low frequency limit 
of frequency-integrals, because the corresponding dependence occuring from 
higher frequencies is clearly regular, and a large $k$-contribution is cut off by the $v_{\bf k}$-factor appearing everywhere in the LLL diagrammatics. Using 
the relations (2.13), a nonanalytic $k$-dependence arising from the lowest 
frequencies is found to be less singular than $k r_B \, {\rm ln}(k r_B)^{-1}$ 
which vanishes in $k \to 0$. We note that, since ${\bf k}$ is always 
accompanied by $r_B$, the pinning strength occuring after a ${\bf k}$-integral 
in a physical quantity is measured by $U_p/r_B^2 \simeq (E_{\rm F} \tau)^{-2}$ 
in the dirty limit in $t < 1$. Since the quantum fluctuation strength 
$U_4/(r_B^2 \gamma)$ becomes, as seen below, of the order $(E_{\rm F} \tau)^{-1}$ in the dirty limit, this implies that 
the relative (effective) pinning strength 
$U_p \gamma/U_4$ is O($(E_{\rm F} \tau)^{-1}$). It is verified in terms of eqs.(2.2) and (2.11) in $t \gg 1$ limit that this is of the same order as 
the thermal counterpart, denoted as $\Delta_{\rm eff}$ in ref.15, 
$\beta U_p(t \gg 1; {\bf k} =0)/U_4$ in the dirty limit. 
Therefore, in the ordinary dirty limit, the vortex pinning strength 
relatively {\it increases} at any $T$ with increasing disorder with strength 
$(E_{\rm F}\tau)^{-1}$.  
 
\section{Quantum VG Transition Point}

In this section, we examine the $T=0$ VG transition point $B_{{\rm vg},0}$ in the Gaussian (i.e., mean field) approximation using the LLL-GL 
action in the dirty limit derived in $\S 2$, and, based on this result, effects of the electron-electron repulsion on $B_{{\rm vg}, 0}$ are qualitatively 
studied. The VG transition field is defined\cite{MPAF,DF,RI1} by examining 
where the uniform VG susceptibility $\chi_{\rm vg}$ (defined below) at zero 
frequency tends to diverge. For a moment (until reaching eq.(3.16)), the GL 
action is assumed to be well-defined in low $T$ limit. 
%The Gaussian approximation predicts not only a $T=0$ 
%transition field $B_{{\rm vg},0}$ of our interest but also a 
%(mean field) transition line $B_{{\rm vg},0}(T)$ at 
%nonzero $T$ with $B_{{\rm vg},0}$ as a $T=0$ end point. 
%Although this line should not be a true transition line in 
%2 D, nevertheless it is expected to be reflected in 
%experimental data of weakly disordered systems 
%as a crossover line at which the resistance remarkably 
%decreases. 

The key quantity $\chi_{\rm vg}$ in this section is proportional to 
$\int_{{\bf r}_1} \int_{{\bf r}_2} {\overline {|<\Psi_{\omega=0}({\bf r}_1) 
\Psi^*_{\omega=0}({\bf r}_2)>|^2}}$ and, consistently with eqs.(2.5) and 
(2.11), is given by 
$$\chi_{\rm vg} = \sum_{p-p'} {\overline {|{\cal G}(p, p', \, 0)|^2}}, 
\eqno(3.1)$$
where ${\cal G}(p,p', \, \omega) = < \varphi_0(p, \omega) \, \varphi^*_0(p', 
\omega)>$ is the LLL fluctuation propagator defined prior to the random 
averaging. On the other hand, the renormalized 2D LLL 
superconducting fluctuation, 
defined through the random-averaged (replica-diagonal) fluctuation propagator
$${\cal G}(\omega) = {\overline {{\cal G}(p,p', \, \omega)}} = ( \, 
\mu(0) + \gamma|\omega| + \Sigma(\omega) \, )^{-1} \eqno(3.2)$$
with self energy $\Sigma(\omega)$, is nondivergent (noncritical) 
even at $T=0$ below the mean field $H_{c2}(0)$, because the dimensionality of 
the LLL fluctuation in 2D is two even at $T=0$. Hence, it erases the {\it mean 
field} superconducting transition on the $H_{c2}(T)$-line to create a first 
order vortex-solidification transition below it.\cite{RI4,BN} In 3D, on the 
other hand, the dimensionality of dissipative LLL fluctuation at $T=0$ becomes three, 
and hence, a "ctitical field" $H_{c2}^*(0)$ can be defined at $T=0$ {\it 
below} the solidification field $B_m(0)$ (see $\S 5$ in ref.8). 

If the (Gaussian) VG transition occurs in a {\it higher} field than 
the disorder-free true transition (i.e., the solidification transition) point, 
a familiar resummation approximation of 
diagrams, such as RPA, is applicable\cite{RI4} in adressing a glass transition
point. Following previous works, we will invoke a systematic loop (or $1/M$) 
expansion and focus on its lowest order ($M=\infty$) terms by, as a mathematical tool, assuming $\Psi$ to have $M$-flavors. It is because 
this approach applied to the 3 D {\it thermal} vortex state 
under a correlated (line-like) disorder parallel to the field has
led\cite{RI1} to a transition 
line qualitatively consistent with the superconducting 
transition in high Tc superconductors with parallel columnar defects. Note
that, since the disorder with strength $U_p$ is, in eq.(3.1) at $T=0$, 
persistently correlated along the ``time'' direction, the present 
quantum GL model with disorder is formally similar\cite{RI1}, close 
to $T=0$, to the bulk 3 D GL model at high temperatures with line-like 
defects parallel to the field. 

Hereafter, we write $\Sigma(\omega) =
\Sigma_{\rm vg}(\omega) + \Sigma_{\rm F}$, where the first and
second terms are, respectively, given by Fig.3 (a) and Fig.3 (b). 
The expression of $\chi_{\rm vg}$ consistent with them has the form of a 
ladder of pinning lines with vertex correction, as described 
in Fig.3 (c) and (d). 
In expressing the diagrams explicitly, the bare pinning strength 
$U_p/(2 \pi r_B^2)$ can be regarded as being replaced by 

$$\Delta^{(R)}_0 = U_p \, \int_{\bf k} v_{\bf k} \, f_{00}(k^2r_B^2) 
\bigg(1 + \sigma_{\rm vg} 
\, v_{\bf k} \bigg)^{-2} = {{U_p} \over {2 \pi r_B^2}} C(\sigma_{\rm vg}), 
\eqno(3.3)$$
where the factor $C(\sigma_{\rm vg}) = \int_0^\infty dk k \, f_{00}(k^2) e^{-k^2/2} (1 + \sigma_{\rm vg} e^{-k^2/2})^{-2}$ with 
$$\sigma_{\rm vg} = \frac{U_4}{2 \pi r_B^2 \beta} \, 
\sum_{\omega''} ( \, {\cal G}({\omega''}) \, )^2 \eqno(3.4)$$
implies a renormalization (vertex correction) due to the LLL fluctuation of the pinning strength. The detailed form of $C(\sigma_{\rm vg})$ depends on the functional form of $f_{00}$ and thus, as mentioned in $\S 2$, is not known 
concretely. We just 
expect $f_{00}({\bf k})$ to have an algebraic form in $k^2$. It implies that, 
when $\sigma_{\rm vg} \gg 1$, $\Delta^{(R)}_0$ will roughly behave like $\sim \sigma_{\rm vg}^{-1} U_p/r_B^2$. The two self energy terms are given by 
$$\Sigma_{\rm vg}(\omega) = - \Delta^{(R)}_0 \, 
{\cal G}(\omega), \eqno(3.5)$$ 
$$\Sigma_{\rm F} =  \frac{U_4}{2 \beta} \sum_{\omega} \int_{\bf k} v_{\bf k} 
{\cal G}(\omega) = \frac{U_4}{4 \pi r_B^2 \beta} \sum_\omega {\cal
G}(\omega), \eqno(3.6)$$
and thus, eq.(3.2) yields  
$$( \, {\cal G}(\omega) \, )^{-1} = \mu(0) + \gamma|\omega|- \Delta^{(R)}_0 
{\cal G}(\omega) + \Sigma_{\rm F}. \eqno(3.7)$$
The Gaussian VG susceptibility $\chi_{\rm
vg}$, consistent with the above $\Sigma_{\rm
vg}(\omega)$, is given as a series of ladder diagrams. Since the
propagator ${\cal G}(\omega)$, due to the LLL degeneracy, 
depends only on a Matsubara frequency
$\omega$, the sum of ladder diagrams is a trivial geometrical series, 
and hence, $\chi_{\rm vg}$ simply becomes 
$$\chi_{\rm vg}= 2 \pi r_B^2 \Delta^{(R)}_0 \bigg( 1 - \Delta^{(R)}_0 
{\cal G}^2(0) \bigg)^{-1} \eqno(3.8)$$
so that the ``Gaussian'' VG transition point is given by the equation 
$$\Delta^{(R)}_0 {\cal G}^2(0) = {{U_p \, C(\sigma_{\rm vg})} \over {2 \pi r_B^2}} {\cal 
G}^2(0) = 1. \eqno(3.9)$$

Eq.(3.7) is quadratic in the renormalized propagator ${\cal
G}(\omega)$ and easily analyzed. Further, at $T=0$ where the frequency 
summation is replaced by an integral, we can accomplish the present analytic 
calculation. Using eq.(3.9), the propagator ${\cal
G}(\omega)$ resulting from eq.(3.7) is expressed at the VG transition field 
$B_{{\rm vg},0}$ by 
$${\cal G}^{-1}(\omega) = {\cal G}^{-1}(0) + {{\gamma|\omega|}
\over 2} + \sqrt{ \gamma|\omega| \biggl({\cal G}^{-1}(0) + {{\gamma|\omega|} \over 4} \biggr)} 
\eqno(3.10)$$
with 
$$2 \, {\cal G}^{-1}(0) = {\rm ln}\biggl({{B_{{\rm vg},0}} \over 
{H_{c2}^d(0)}}\biggr) + \Sigma_F, \eqno(3.11)$$
where eq.(3.9) was used. 
The $\sigma_{\rm vg}$-expression at $B_{{\rm vg},0}$ is expressed as 
$$\sigma_{\rm vg}={{U_4} \over {4 \pi^2 r_B^2}} \int_0^\infty d\omega 
{\cal G}^2(\omega) = {{U_4} \over {6 \pi^2 \gamma r_B^2}} {\cal G}(0), 
\eqno(3.12)$$
and, by combining this with eq.(3.9), $\sigma_{\rm vg}$ is a function of the combination 
$$\eta = {{U_4^2} \over {18 \pi^3 \gamma^2 r_B^2 U_p}} \eqno(3.13)$$
and given as a solution of 
$$\sigma_{\rm vg}^2 \, C(\sigma_{\rm vg}) = \eta. \eqno(3.14)$$ 
On the other hand, the integral of $\Sigma_F$ requires a high-frequency
cut off. By reasonably assuming a constant of order unity $\Lambda_c \sim
\gamma \omega_M$, where $\omega_M$ is a frequency cut off, the frequency sum 
(integral) of eq.(3.6) results in 
$$\Sigma_F = {{U_4} \over {4 \pi^2 r_B^2 \gamma}} \, 
{\rm ln} \biggl({{6 \pi^2 \gamma r_B^2 \sigma_{\rm vg} \Lambda_c} 
\over {U_4}} \biggr). \eqno(3.15)$$
Applying this to eq.(3.11), we obtain the relation 
$${\rm ln}\biggl({{B_{{\rm vg},0}} \over {H_{c2}^d(0)}}\biggr) 
= {{U_4} \over {4 \pi^2 r_B^2 \gamma}} \biggl({4 \over {3 
\sigma_{\rm vg}}} - \, {\rm ln} \biggl({{6 \pi^2 \gamma r_B^2 
\sigma_{\rm vg} \Lambda_c} \over {U_4}} \biggr) \, \biggr). \eqno(3.16)$$
Eqs.(3.13), (3.14), and (3.16) give the $T=0$ VG transition field $B_{{\rm vg}, 0}$. In the bracket of r.h.s. of eq.(3.16), the first term is a measure of the 
vortex pinning strength, while the second term is a measure of quantum LLL 
fluctuation effect. 

First, let us apply the above result to the ordinary dirty limit. 
According to eqs.(2.2) and (2.4), the coefficients in GL action 
take the following values in $T \to 0$ limit 
$$\gamma \to \gamma^{(0)}(T=0) = {{\tau \phi_0} \over {\pi B \, l^2}}, $$
$$U_4 \to N^{-1}(0) \, (\gamma^{(0)}(T=0))^2, \eqno(3.17)$$ 
together with eq.(2.10). Consequently, in this case, $\eta$ is a universal number, $5 \times 10^{-3}$, and hence, $\sigma_{\rm vg}$ is also a constant, possibly of the order of $10^{-1}$. Further, the strength $U_4/(4\pi^2 r_B^2 \gamma)$ of quantum superconducting fluctuation becomes $(2 \pi E_{\rm F} \tau)^{-1}$. 
Then, eq.(3.16) implies that, although $B_{{\rm vg},0}$ will lie below 
$H_{c2}^d(0)$ for large enough $E_{\rm F} \tau$, it {\it increases} 
with increasing 
the disorder strength $1/(E_{\rm F} \tau)$. One can find 
that the primary origin 
of this $B_{{\rm vg},0}$-increase is the disorder-induced enhancement 
of $H_{c2}^d(0) \propto \tau^{-1}$, while the cancellation between the two 
terms in the bracket of eq.(3.16) is subtle and may depend on the 
diagram-resummation 
method which should be refined in higher orders of $(E_{\rm F} \tau)^{-1}$. 
Here we will respect the present result in the simple but {\it consistent} 
$M^{-1}=0$ approximation and expect, more generally (but in the dirty limit), the fluctuation contribution to outweigh the pinning contribution and to 
make $B_{{\rm vg},0}$ lower than $H_{c2}^d(0)$ at least 
up to O($(E_{\rm F} \tau)^{-1}$) (see also the next paragraph). 
On the other hand, {\it without} the fluctuation (second) term in eq.(3.16), 
it would show a $T=0$ 
superconducting transition point increasing with increasing disorder and 
existing {\it above} $H_{c2}^d(0)$. 
This statement is essentially the same as the argument by Spivak and Zhou\cite{Spiv} of an unlimitedly large "$H_{c2}$" (see 
also the sentence prior to eq.(2.7)). As already mentioned, we expect the 
inclusion of the quantum superconducting fluctuation to push the 
superconducting transition point down to $B_{{\rm vg},0}$ below $H_{c2}^d(0)$, 
even in the ordinary dirty limit. 

The true transition point $B_{\rm vg}$ should be lowered further 
from the Gaussian one $B_{{\rm vg},0}$ by going beyond the present 
Gaussian approximation and taking account of interactions between the VG 
fluctuations. Explaining this requires another apparatus and will be given in 
a separate paper\cite{RI5}. We note here that this shift $1- B_{\rm vg}
/B_{{\rm vg},0}$ is also of the order of $(E_{\rm F} \tau)^{-1}$. 
Nevertheless, this fact does not change the above statement in the ordinary 
dirty limit that $B_{\rm vg} < H_{c2}^d(0)$, while $B_{\rm vg}$ will increase 
with increasing $(E_{\rm F} \tau)^{-1}$. Below, we will not distinguish 
$B_{\rm vg}$ from $B_{{\rm vg},0}$ in discussing effects of an 
electron-repulsion. This simplification does not affect conclusions 
which follow in this section. 

Now, let us examine how an interplay between an electron-repulsion and 
disorder changes the above results in the ordinary dirty limit. According to 
the previous works\cite{Fin} (see also ref.19), the contributions of the 
dynamically-screened Coulomb interaction to the linearized GL equation for 
quasi 2D films can be included perturbatively by assuming in 2D case a short-ranged repulsive interaction with strength $\lambda_1= R_r/(8 \pi R_Q) = 3/(k^2_{\rm F} \, d \, l)$, where $d$ is the film thickness, and $R_r$ is identified with the high temperature sheet resistance. As shown in ref.9, the corrections due 
to the short-ranged electron repulsion to the GL coefficients $\mu(0)$ and 
$U_4$ are convergent in low $T$ limit at each order in $\lambda_1$, since the 
frequency dependences of Cooperons appearing as vertex corrections for the 
couplings to the pair-fields are cut off by the $B$-dependence. It implies 
that $\mu(0)$ and $U_4$ are determined by the high frequency side of Matsubara 
frequency summations and that their $T$ dependences will, irrespective of 
$\lambda_1$-values, be lost roughly when $t < 1$, i.e., $T < T_{cr}^{\rm mf} 
\equiv 0.15 T_{c0} B/H_{c2}^d(0)$ (see $\S 2$). 
As mentioned above eq.(2.7), the situation is also 
similar in $\lambda_1$-dependences of $U_p f_{00}({\bf k}r_B)$. 
In contrast, each term of $\lambda_1$-perturbation series for the time 
scale $\gamma$ is logarithmically divergent, and, at low enough $T$, $\gamma$ 
has the systematic expansion parameter\cite{RI3,com} 
$\lambda_1 {\rm ln}(T/T_{cr}^{\rm mf})$. It is expected from this systematic 
perturbation series that an onset temperature below which $\gamma$ begins to 
rapidly decrease on cooling will be given by 
$$T_{\rm rep}(\lambda_1) \simeq T_{cr}^{\rm mf} 
\exp(-c_{\rm rep}/\lambda_1),  \eqno(3.18)$$ 
where $c_{\rm rep}$ is a positive constant (possibly) slightly less than 
unity. These facts imply that we have an intermediate temperature region 
below $T_{cr}^{\rm mf}$ but above $T_{\rm rep}(\lambda_1)$. Besides 
these {\it microscopic} temperature scales, we have the quantum-thermal 
crossover temperature $T_{cr} \equiv U_4/(2 \pi r_B^2 \gamma^2)$ on the LLL fluctuation behavior (see $\S 2$ in ref.8). Then, 
if $T_{\rm rep} \ll T_{cr}$ ($< T_{cr}^{\rm mf}$), there is a low but 
intermediate temperature region 
above $T_{\rm rep}$ but below $T_{cr}$ in which any microscopic 
$T$-dependence carried by 
the GL coefficients is negligible and the LLL fluctuation is of quantum 
nature, although the GL coefficients are different from those in the dirty 
limit due to the $\lambda_1$ corrections. In $\S 4$, a computational evidence on the presence of such an intermediate temperature range will be given. 
Then, an {\it apparent} VG transition field 
$B_{\rm vg}^*$ can be {\it defined} in this intermediate temperature region as 
a "$T=0$" transition field and, according to eq.(3.16), is expressed by 
$$B_{\rm vg}^* \simeq H_{c2}(0) (1 - d_{\rm g}^*(E_{\rm F}\tau)^{-1} \, ), \eqno(3.19)$$ 
up to O($(E_{\rm F} \tau)^{-1}$),
where $H_{c2}(0)$ and the constant $d_{\rm g}^*$ have 
$\lambda_1$-corrections. Although we expect $d^*_{\rm g}$ to be positive, it is possibly smaller than unity, at least when $\lambda_1=0$, once 
recalling the above-mentioned cancellation between a fluctuation term and a 
pinning term. Although the $\lambda_1$-correction to $d_{\rm g}^*$ 
is due to those in $U_4$ and $U_p$, it is not easy to exactly obtain those corrections even up to O($\lambda_1$). However, since the $\lambda_1$-correction in $d_{\rm g}^*$ is accompanied in eq.(3.19) by the factor $1/E_{\rm F}\tau$, 
the $\lambda_1$-dependence of $B_{\rm vg}^*$ can be seen as being 
dominated by that of $H_{c2}(0)$. It is now known that $H_{c2}(0)$ 
decreases\cite{RI3,Smith} with increasing 
$R_r$, and thus, the {\it apparent} critical field $B_{\rm vg}^*$, defined in 
the intermediate temperature range, is expected to decrease with increasing 
$R_r$, as indicated in experiments.\cite{ex1,ex2,ex3} 

To find the $R_r$-dependence of a {\it true} (but, possibly, inaccessible) 
$B_{\rm vg}$ at $T=0$, we need the 
$\lambda_1$-dependences of the GL coefficients in low $T$ limit 
$T \ll T_{\rm rep}$. 
Since, at the present stage, we have no computation evidence enough to argue 
that $\gamma$ will remain positive in low $T$ limit, we will merely assume here $\gamma(T \to 0)$ to approach a small but positive value $\gamma_{\rm min}$. 
Then, according to 
eq.(3.16), $B_{\rm vg}$ is given by eq.(3.19) with $d^*_{\rm g}$ there 
replaced by $d_{\rm g}$, which will take the form  
$$d_{\rm g} \simeq {{\gamma^{(0)}} \over {2 \pi \, \gamma_{\rm min}}} \, 
{\rm ln}\biggl({{\gamma^{(0)}} \over {\gamma_{\rm min}}} \biggr) 
\gg d_{\rm g}^*, \eqno(3.20)$$ 
where other $\gamma_{\rm min}$-independent terms were neglected by assuming $\gamma_{\rm min} \ll \gamma^{(0)}$. Since it will be clear that $\gamma_{\rm min}$ is insensitive to $\lambda_1$ or decreases with increasing $\lambda_1$, 
$B_{\rm vg}$ will decrease, more drastically than $B_{\rm vg}^*$, with 
increasing $R_r$. However, we believe through the results in $\S 4$ and 
experimental informations that the field $B_c$ at which a flat ($T$-independent) resistance curve is seen should correspond to $B_{\rm vg}^*$ and that 
$B_{\rm vg}$ is not measurable because $T_{\rm rep}$ is inaccessibly low.  

Finally, let us compare eq.(3.19) with the corresponding expression on 
$B_m(0)$. By substituting the parameter values in the dirty limit into the expression\cite{com2} $|\mu_0(0)| \propto U_4^{(0)} /(\pi r_B^2 \gamma^{(0)})$ 
derived in ref.8 (see $\S 2.2$ there), we obtain 
$$B_m(0) \simeq H_{c2}(0) (1 - d_m (E_{\rm F} \tau)^{-1}), \eqno(3.21)$$
where the constant $d_m$ was argued there\cite{RI2} to be more than 6.0. 
By taking account of the prospects on $d_{\rm g}^*$-value mentioned above, 
we believe here that, in general, $B_m(0)$ will lie below $B_{\rm vg}^*$. Further, due to the similarity on parameter dependences in eqs.(3.19) and (3.21), 
the difference between $B_m(0)$ and $B_{\rm vg}^*$ should enlarge with 
increasing $R_r$. 

\section{Computation of $\gamma$ at low $T$}

As mentioned in $\S 3$, we have previously\cite{RI3} shown that $\Gamma(T) 
\equiv \gamma(T)/\gamma^{(0)}(T=0)$ at low enough $T$ takes a form of power 
series\cite{com} in $\lambda_1 {\rm ln}(T/T_{cr}^{\rm mf})$ suggestive of the 
presence of a temperature scale $T_{\rm rep}(\lambda_1)$ of the form (3.18), 
where $\gamma^{(0)}(T=0)$ is the limiting value given in eq.(3.17). 
To demonstrate our argument in $\S 3$ on the presence of the $T$-insensitive 
intermediate region above $T_{\rm rep}$, we have carried out a numerical calculation of $\Gamma(T)$ useful even at low enough $T$ on the basis of 
the resummation technique of Oreg and Finkel'stein (OF)\cite{Fin2}. Its 
preliminary results will be briefly reported here. 

The OF's technique was originally developed to obtain the reduction of 
mean field $T_c$ in low dimensional $s$-wave case 
due to the interplay between the electron-repulsion and disorder and 
subsequently, was extended 
to nonzero field case in ref.19 to examine $H_{c2}(T)$-lines. 
This method focuses on the selfconsistent equation for the vertex 
part ${\hat V}_c$ in the Cooper 
channel and consists of solving it as a matrix equation in the (Matsubara) 
frequency space. The mean field transition point is obtained as a vanishing 
eigenvalue $E_c(\omega=0)$ of the inverse of ${\hat V}_c(\omega=0)$, where 
$\omega$ is the Matsubara frequency carried by the pair-field. One can manage 
to, in order to find $\Gamma$ valid beyond the lowest order in $\lambda_1$, 
extend this technique to the case with nonzero $\omega$ by applying\cite{Serene} a Pade-approximant to the frequency dependence of $E_c$. 

In Fig4, the $\Gamma$ v.s. $T$ curves computed for various $\lambda_1$ 
values and at the field $H_{c2}(0)$ of each $\lambda_1$ are given. In the temperature range $T/T_{c0} \leq 0.15$ (i.e., $T < T_{cr}^{\rm mf}$) of our interest, the temperature variations of $\Gamma$ seem to become weaker with increasing 
$\lambda_1$ except at the lowest temperatures. The reduction of 
$\gamma$-values accompanying the $\lambda_1$-increase at the intermediate 
temperatures arises, as well as that of $H_{c2}$-values, from the high 
frequency contribution in the diffusion propagators which was neglected for brevity in the $\gamma$-expression calculated in ref.9. It is not easy to judge how $T_{\rm rep}$ should be defined from the curves, and it will be defined, 
for convenience, as the temperature below which $\Gamma$ becomes less than 
$\Gamma(T/T_{c0}=0.15)$. Then, we have, for instance, 
$T_{\rm rep}(R_r/R_Q = 0.257) \simeq 0.005 T_{c0}$, and $T_{\rm rep}(R_r/R_Q 
= 0.428) \simeq 0.013 T_{c0}$. Further, according to Fig.5 in which $R_r/R_Q = 0.428$ is commonly used, the $T_{\rm rep}$ thus defined decreases with 
decreasing $B$ consistently with eq.(3.18) proportional to $T_{cr}^{\rm mf}$. 
Based on these figures, we conclude that the presence of an intermediate 
temperature region, in which $\gamma$ and hence, $B_{\rm vg}^*$ can be defined 
as quantities insensitive to $T$, has been justified by the above microscopic 
computation. 

\section{Summary and Discussion}

In this paper, we have examined the position and parameter dependence of 2D 
VG transition field to be expected in low $T$ limit. In the realistic model with repulsive interaction between electrons, we have a complicated situation due 
to a {\it microscopic} $T$-dependence 
of the time scale $\gamma$ becoming remarkable rather {\it below} a very low temperature scale $T_{\rm rep}$: An intermediate (but low) 
temperature range above $T_{\rm rep}$, rather than the low $T$ limit below it, 
is expected to be relevant to 
experiments suggesting a 2D FSI behavior so that an {\it apparent} critical 
field $B_{\rm vg}^*$, being insensitive to $T$ there, plays the role of 
a "$T=0$" VG critical field. In a companion paper,\cite{RI5} we will show how 
available resistivity data suggesting a $T=0$ FSI transition can be explained 
based on the present results and argument. 

Finally, we wish to connect the present result with the resistive behavior near the {\it disorder free} 2D quantum melting line\cite{RI2} $B_m(T) \simeq 
B_m(0)$. In ref.8, it was pointed out that, at nonzero temperatures 
in the quantum regime $T < T_{cr}$, the usual 
fluctuation conductance results in "fan-shaped" resistivity curves similar to 
the 2D FSI behavior near or below a fluctuation-corrected $H_{c2}(0)$ and, only close to and below $B_m$, reduces to the classical vortex flow 
behavior at the same $T$. 
Further, it was argued even that this itself may be the origin of the 2D 
FSI behavior. This argument based on the neglect of vortex pinnings 
may be justified only if $B_m > B_{\rm vg}^*$. However, as seen at the end of 
$\S 3$, $B_m(0)$ is in general likely to lie below $B_{\rm vg}^*$. Therefore, 
to try to understand comprehensively the FSI behaviors in disordered thin 
films, a VG fluctuation contribution to conductance needs to be included which 
will be examined in ref.17. In fact, recent data analysis\cite{ex1} has 
suggested a vortex lattice melting to occur much 
below the critical field $B_c$ at which the resistance 
becomes flat (see ref.17). We wish to note that, nevertheless, 
the knowledges\cite{RI2} on the disorder-free 
fluctuation conductance at low $T$ become important\cite{RI5} in understanding 
differences in the intervening metallic resistance value at $B_{\rm vg}^*$ between various materials. 

After submitting this manuscript, we were aware of the paper by Galitski and 
Larkin\cite{GL} who have also examined a quantum transition field on a 
macroscopic superconductivity in disordered 
thin films from a different point of view and by neglecting an 
electron-electron repulsion. Although, in contrast to our approach, they have 
assumed a spontaneous creation of granular structure, 
nevertheless an expression determining the transition field at $T=0$ has been 
derived which is essentially the same as our eq.(3.16), except for the absence in their expression of eq.(3.16)'s second term (note that $E_{\rm F} \tau$ is 
denoted as $g$ in ref.23), and has led them to a conclusion similar to our 
statement given below eq.(3.17) in relation to ref.14. However, the quantum amplitude fluctuation and the microscopic interplay between disorder and an electron-electron repulsion, both of which should contribute to a decrease of a quantum transition field, have been ignored in ref.23. Further, although a glass behavior is expected in lower fields at $T=0$, the transition field was determined 
there, according to the sentences below their eq.(16), by assuming an 
occurrence of the {\it ordinary} phase coherence in 
contradiction to our theory with the amplitude fluctuation included showing 
that the glass ordering is not signaled by a development of the ordinary phase 
coherence. Explanation of experimental data listed 
in ref.23 should be ascribed to thermal fluctuation effects\cite{ROT,RI4}. Details of this discussion will be given elsewhere\cite{PRL}. 

\vspace{5mm}

\leftline{\bf Acknowledgement}

Numerical computation in this work was carried out at the Yukawa Institute 
Computer Facility in Kyoto University.

\vfil\eject

\leftline{Figure Captions}

\vspace{10mm}
Fig.1 

Feynman diagrams contributing to ${\cal F}$ and belonging to the same family 
when classified according to how the diffusion propagators appear. Double solid lines imply the pair-field propagators, a single solid line is an electron propagator, the dotted line with open circle is a diffusion propagator between different replicas, 
the dotted line with cross denotes 
a single impurity line carrying $(2 \pi N(0) \tau)^{-1}$, and the hatched 
corner vertex parts imply the Cooperons which modify the couplings to the 
pair-fields. 

\vspace{5mm}

Fig.2

Details of the first three diagrams in Fig.1. 

\vspace{5mm}

Fig.3 

Diagrams necessary in obtaining $B_{{\rm vg},0}$, where a hatched corner 
vertex is defined in (c) (which should not be confused with that in Fig.1 and 2), the double dotted line denotes the pinning line carrying $U_p$, and the solid circle is the interaction strength carrying $U_4$. See the text for other 
details. 

\vspace{5mm}

Fig.4 

$\Gamma(T)$ v.s. $T$ curves computed in terms of OF's 
technique for different values of $8 \pi \lambda_1$ $= R_r/R_Q = 0$ (top), 
$0.086$, $0.171$, $0.257$, $0.342$, and $0.428$ (bottom). 
Each curve was obtained by fixing $B$ to the $H_{c2}(0)$-value at each 
$\lambda_1$-value, and $2 \pi T_{c0} \tau = 0.25$ was used commonly. 

\vspace{5mm}

Fig.5 

$\Gamma(T)$ v.s. $T$ curves obtained for 
different field values, $B/H_{c2}(0)$ $=1.2$ (top), $1.0$, and $0.8$ (bottom) by fixing $8 \pi \lambda_1$ and $2 \pi T_{c0} \tau$ to $0.428$ and $0.25$, respectively. 

\begin{figure}[t]
\begin{center}
\leavevmode
\epsfysize=15cm
\epsfbox{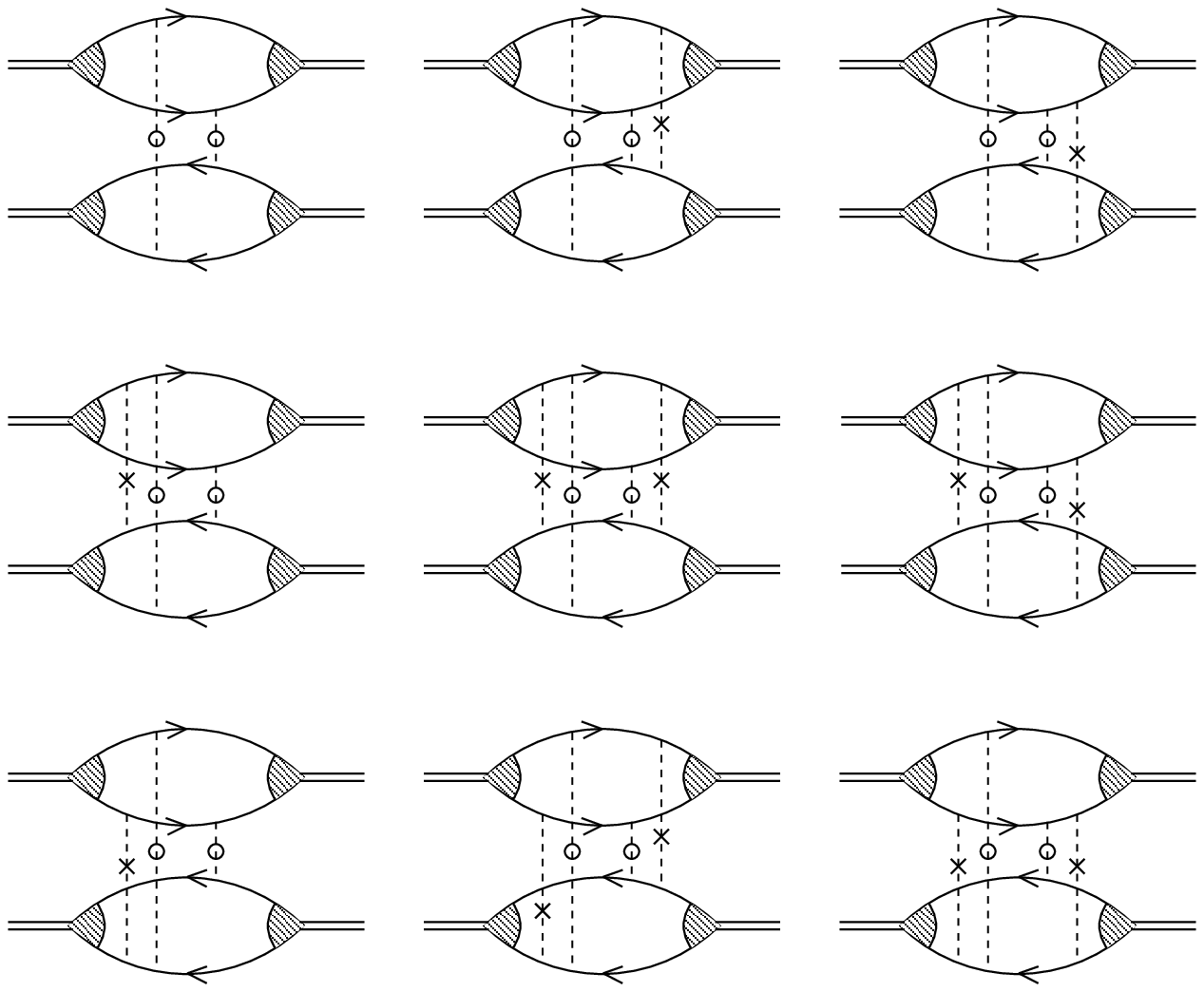}
\end{center}
\end{figure}

\begin{figure}[t]
\begin{center}
\leavevmode
\epsfysize=12cm
\epsfbox{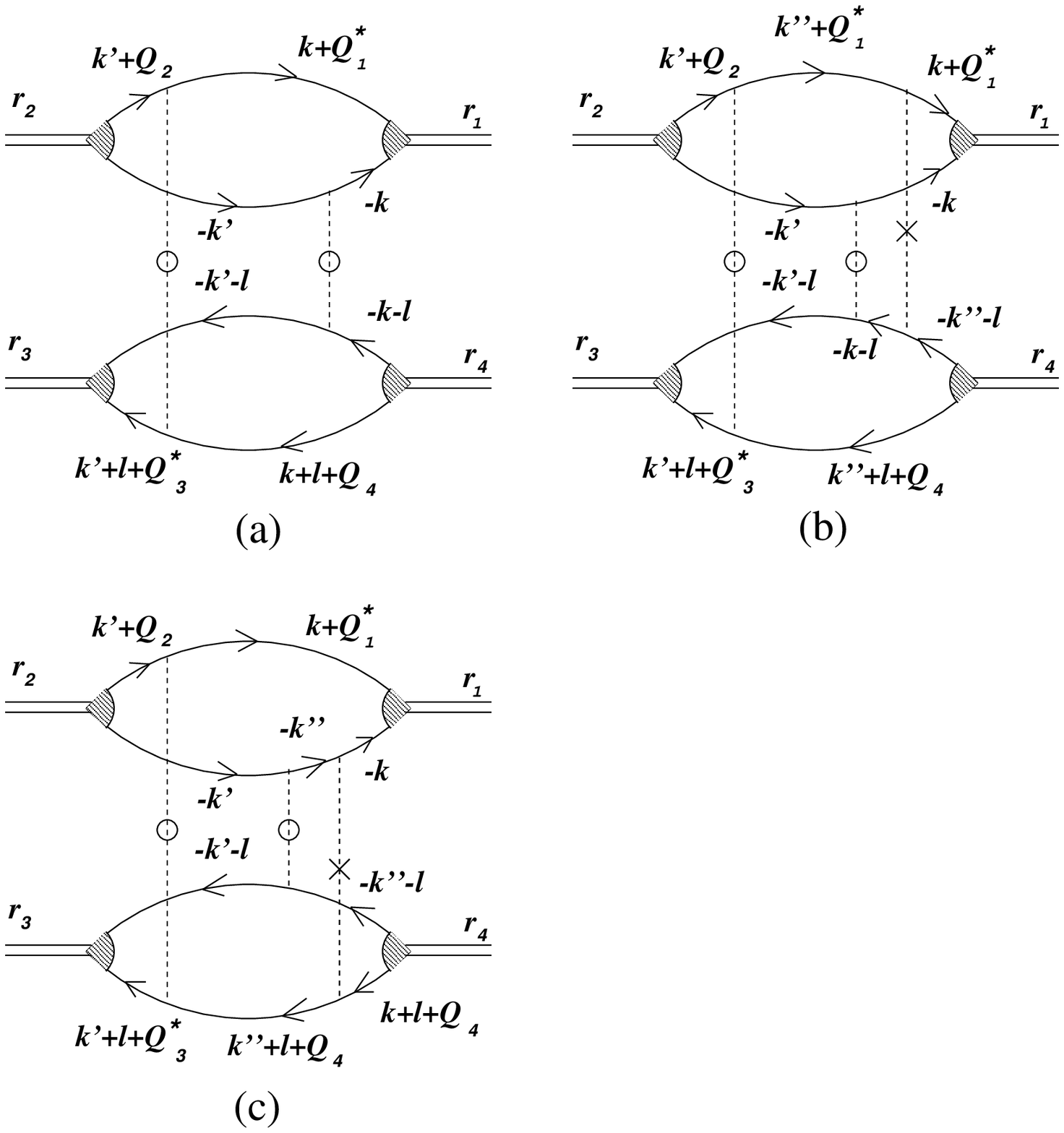}
\end{center}
\end{figure}

\begin{figure}[t]
\begin{center}
\leavevmode
\epsfysize=12cm
\epsfbox{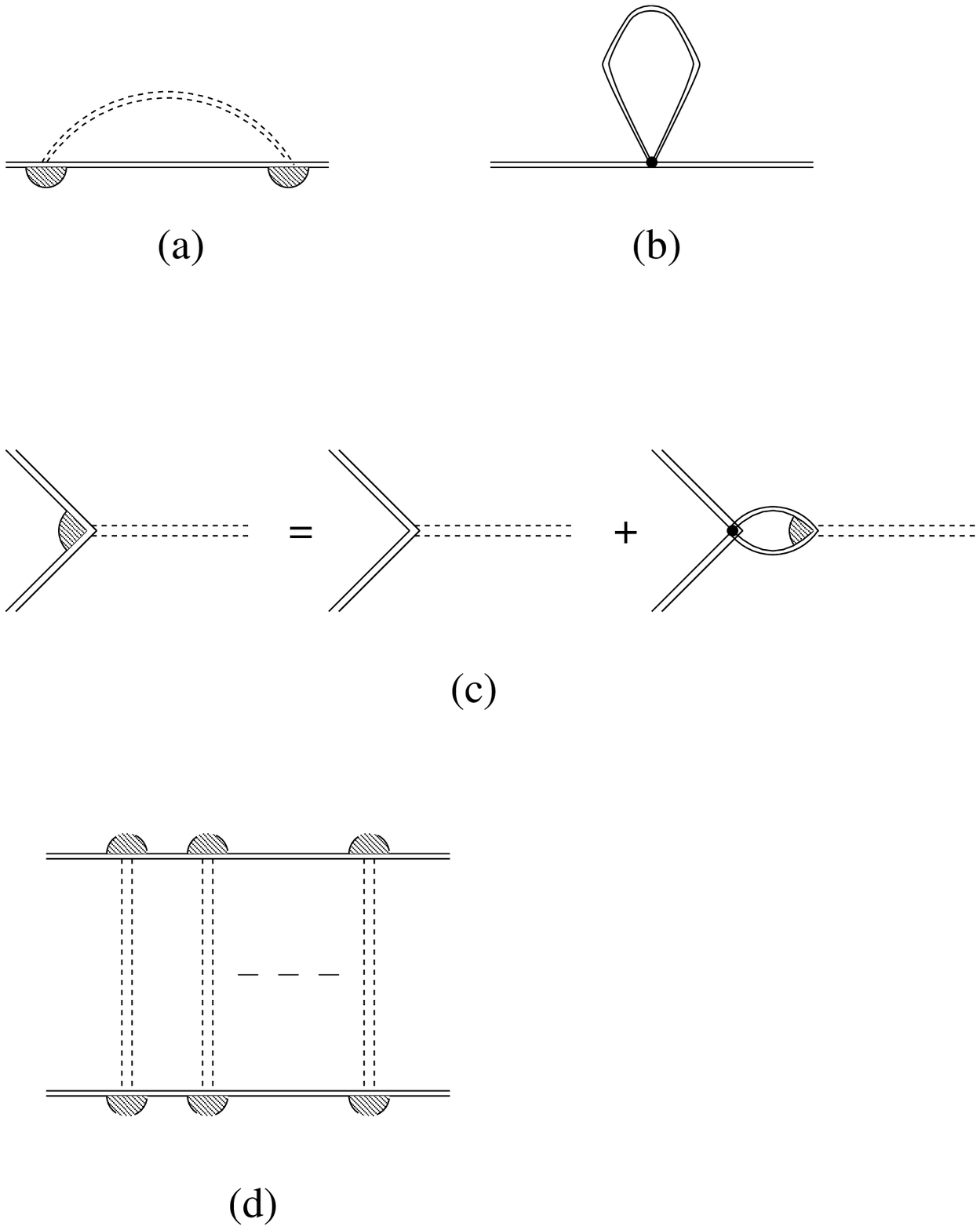}
\end{center}
\end{figure}

\begin{figure}[t]
\begin{center}
\leavevmode
\epsfysize=15cm
\epsfbox{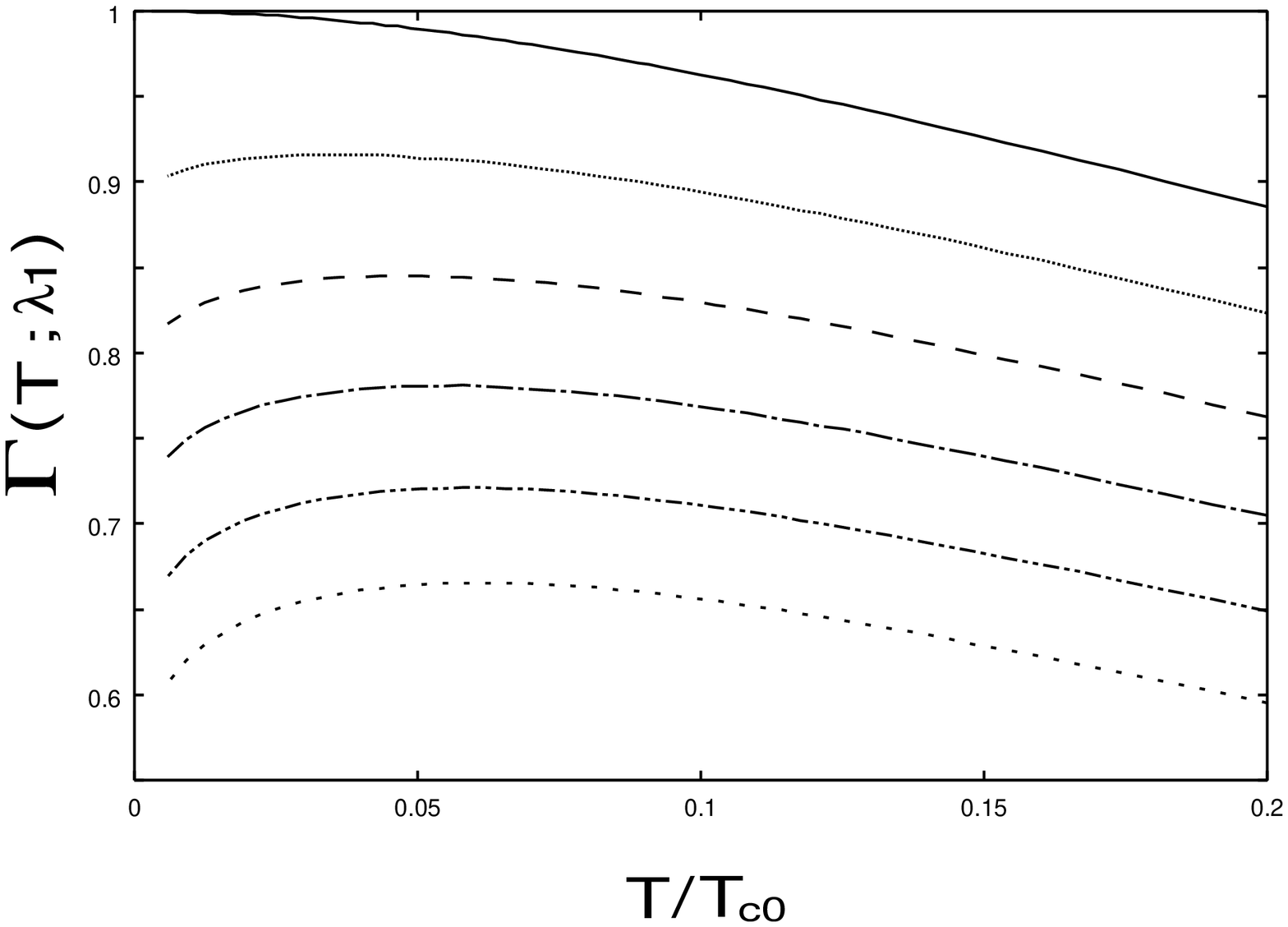}
\end{center}
\end{figure}

\begin{figure}[t]
\begin{center}
\leavevmode
\epsfysize=10cm
\epsfbox{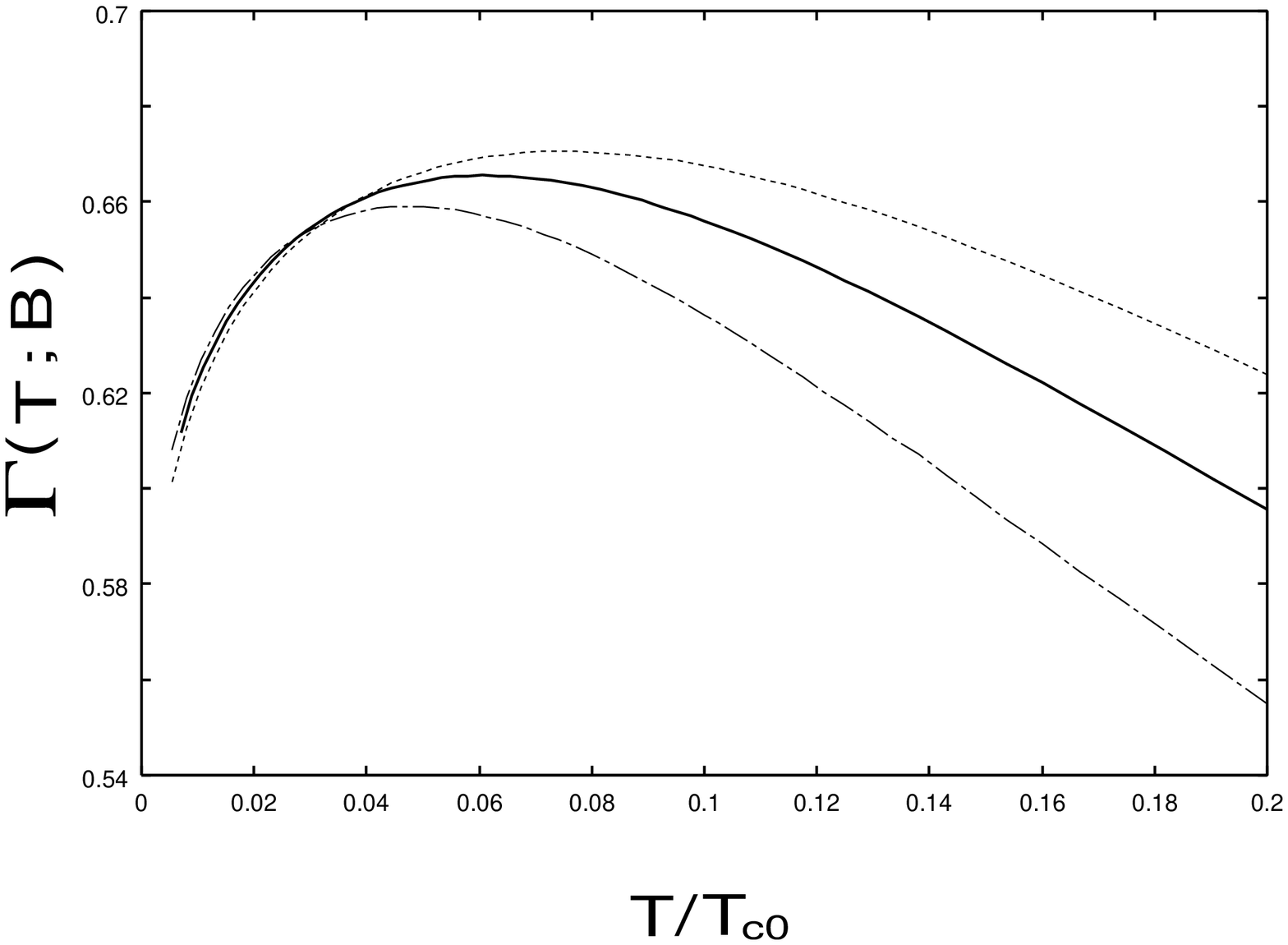}
\end{center}
\end{figure}


\begin{thebibliography}{99}
\bibitem{com} In writing down eqs.(2.20) and (4.9) in ref.9, the fact that the frequency values larger than $T_{c0} B/H_{c2}^d(0)$ do not contribute in obtaining them was ignored. Once modifying this point, $T_{cr}^{\rm mf}$ in place of $(\pi \tau)^{-1}$ plays the role of upper cut-off of the logarithm in the 
expansion parameter for $\gamma$.
\end{thebibliography}
\end{document}